\documentclass[showpacs,amsmath,
twocolumn,
aps,prb]{revtex4}
\usepackage{graphicx}
\usepackage{dcolumn}

\begin{document}

\title{High-field optically detected nuclear magnetic resonance in GaAs}
\author{M. Poggio}
\author{D. D. Awschalom}
\affiliation{Center for Spintronics and Quantum Computation, 
University of California, Santa Barbara, CA 93106}
\date{\today}

\begin{abstract}
  A method for high-field optically detected nuclear magnetic
  resonance (ODNMR) is developed sensitive to $10^{8}$ nuclei.
  Nuclear spin transitions are induced using a radio frequency coil
  and detected through Faraday rotation spectroscopy. Unlike
  conventional ODNMR, which is limited to low fields and relies on the
  measurement of time-averaged luminescence polarization, this technique
  monitors nuclear polarization through time-resolved measurements of
  electron spin dynamics. Measurements in a (110) GaAs quantum well
  reveal $^{69}$Ga, $^{71}$Ga, and $^{75}$As resonances and their
  quadrupolar splittings while resolving changes in nuclear
  polarization of 0.02\%.
\end{abstract}

\maketitle

The small number of nuclear spins in quantum wells (QWs) and quantum
dots makes conventional nuclear magnetic resonance (NMR) experiments
difficult in these semiconductor nanostructures. Optical pumping
\cite{Lampel:1968} strongly enhances nuclear spin polarization and can
increase the detection sensitivity of typical radio frequency (RF)
probes from a minimum of $10^{17}$ nuclear spins to $10^{12}$.  As a
result, RF detection of optically pumped GaAs multiple QWs has been
achieved \cite{Barrett:1994}. Detection of NMR has also been
demonstrated through optical measurements of recombination
polarization, either by exciting NMR transitions with a conventional
coil \cite{Ekimov:1972,Paget:1977}, or by purely optical means
\cite{Kalevich:1980,Kalevich:1981,Kalevich:1986,Eickhoff:2002}. In the
latter case, an optical field is modulated at the nuclear Larmor
frequency resulting in an oscillating electron magnetization.  This
magnetization interacts with nuclear spins through the contact
hyperfine coupling and induces NMR transitions in lieu of an external
RF field. While optically detected NMR (ODNMR) provides the high
sensitivity typical of optical techniques, it has several limitations.
For electron g-factors and spin lifetimes typical of GaAs structures,
ODNMR is only possible at low magnetic fields ($< 1$ T). In addition,
the reliance on radiative recombination for detection makes ODNMR
disproportionately sensitive to nuclei located near shallow donors and
impurities \cite{Paget:1982}.

Another type of ODNMR is possible using time-resolved Faraday rotation
(FR) \cite{Crooker:1997} to probe nuclear spin polarization. In this
detection scheme, FR measures the spin precession frequency of
electrons in the conduction band. Nuclear spins act on electron spins
through the contact hyperfine interaction altering this frequency and
allowing for the precise measurement of nuclear polarization.
All-optical versions of this method have been demonstrated in bulk
GaAs and in a GaAs QW \cite{Kikkawa:2000,SalisPRL:2001,SalisPRB:2001}.
These measurements can be made at high applied magnetic fields and,
unlike measurements of time- and polarization-resolved
photoluminescence, they are not limited by the charge recombination
time.

Here we present an extension of this technique utilizing an RF coil
for the excitation of NMR transitions. The use of an external RF field
allows for the future application of well-developed pulsed NMR
techniques for noise reduction while at the same time exploiting the
high sensitivity of FR detection. In addition, the conventional
magnetic excitation of nuclear transitions circumvents the complex
interactions between electrons and nuclei which take place in optical
excitation schemes. Unlike conventional RF magnetic fields, which
induce only dipole transitions, modulated optical fields induce both
magnetic dipole transitions and electric quadrupole transitions
\cite{SalisPRL:2001,SalisPRB:2001,Eickhoff:2002,Poggio:2003}.

\begin{figure}[b]\includegraphics{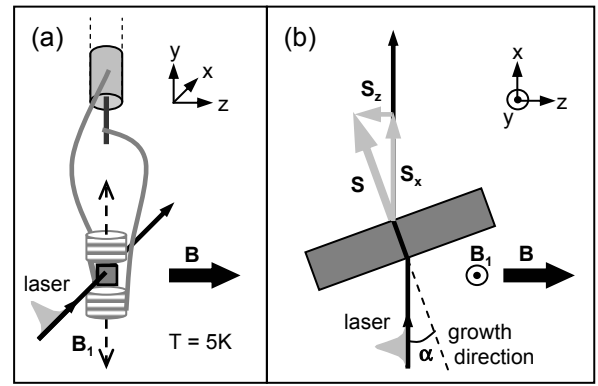}\caption{\label{fig1} 
    Schematic of the experimental geometry. (a) Side view of the
    apparatus within a magneto-optical cryostat. (b) Top view of the
    sample at the center of the Helmholtz coil.  }\end{figure}

As shown in Fig. \ref{fig1}, a semiconductor sample is cooled to $ T =
5$ K in a magneto-optical cryostat with an applied magnetic field
$\vec{B_0}$ along the z-axis and is mounted in the center of a 10 mm
$\times$ 5 mm Helmholtz coil wound from 22 AWG magnet wire. RF
radiation is coupled to the coil from the top of the cryostat through
an impedence-matched semi-rigid coaxial transmission line producing an
RF magnetic field $\vec{B_1}$ along the y-axis. The sample growth
direction lies in the xz-plane and can be rotated to adjust the angle
$\alpha$ between the growth direction and the laser propagation
direction along the x-axis. Unless otherwise specified $\alpha =
10^{\circ}$.

We measure time resolved FR in a modulation doped 7.5-nm wide (110)
GaAs QW with a mobility of 1700 cm$^2$ V$^{-1}$ s$^{-1}$ and and
electron density of $9 \times 10^{10}$ cm$^{-2}$ at $ T = 300$S K.
Confinement along the (110) crystal direction suppresses
D'yakonov-Perel spin scattering resulting in spin lifetimes longer
than 1 ns from $T = 5$ K to room temperature \cite{Ohno:1999}. A
250-fs 76-MHz Ti:Sapphire laser tuned near the exciton absorption
energy (1.572 eV) produces pulses which are split into pump and probe
with a full width at half maximum (FWHM) of 8 meV and an average power
of 2.0 mW and 100 $\mu $W respectively.  The linearly (circularly)
polarized probe (pump) is modulated by an optical chopper at $f_{1} =
940$ Hz ($f_{2} = 3.12$ kHz). Both beams are focused on the sample
surface to an overlapping spot 50 $\mu$m in diameter with the pump
beam injecting polarized electron spins along the sample growth
direction as shown in Fig.  \ref{fig1}b.  The pinning of the of the
initial electron spin polarization $\vec{S}$ along the growth
direction relies on the fact that pump pulses couple predominantly to
heavy hole states, which are split off from light holes states in a QW
\cite{SalisPRB:2001,Marie:2000}. Small rotations in the linear
polarization of the transmitted probe are measured and are
proportional to the component of electron spin polarization in the
conduction band along the growth direction. Variation of the
pump-probe time delay $\Delta t$ reveals the time evolution of this
spin polarization. In the absence of nuclear polarization, electron
spins precess about an axis and at a frequency defined by the Larmor
precession vector $\vec{\nu _{L}} = \hat{g} \vec{B_{0}} \mu_B /h$,
where $\hat{g}$ is the Land\'{e} g-factor expressed as a tensor,
$\mu_{B}$ is the Bohr magneton, and $h$ is Planck's constant. GaAs QWs
grown in the (110) direction exhibit strong anisotropy in $\hat{g}$
resulting in both the dependence of $\nu_L$ on the orientation of
$\vec{B_0}$ with respect to the sample's crystal axes and in a
difference between the precession axis $\vec{\nu_L}$ and the direction
of $\vec{B_0}$ \cite{SalisPRB:2001}.

At $T = 5$ K, spin-polarized photo-excited electrons generate nuclear spin
polarization within the QW through dynamic nuclear polarization (DNP)
\cite{Lampel:1968}.  DNP acts through the contact hyperfine
interaction, written as $A_H \vec{I} \cdot \vec{S} = \frac{1}{2} A_H
\left (I^+ S^- + I^- S^+ \right ) + A_H I_z S_z$, where $A_H$ is the
hyperfine constant and $\vec{I}$ is the nuclear spin. This
``flip-flop'' process results in an average nuclear spin $\langle
\vec{I} \rangle$ along $\vec{B_0}$ and is driven by the component of
electron spin $S_z$ in that direction. The sign and magnitude of
$\langle \vec{I} \rangle$ depends on the angle $\alpha$.

The presence of a non-zero $\langle \vec{I} \rangle$ in turn acts on
the electron spin dynamics through the addition of a term to the
precession vector $\vec{\nu_L} = \hat{g} \vec{B_0} \mu_B / h + A_H
\langle \vec{I} \rangle / h$. The measurement of $\nu_L$ and the
knowledge of $\hat{g}$ and $\vec{B_0}$ yield the nuclear polarization
frequency $\nu_n = A_H \langle I \rangle / h$, which has been
calculated to be 32.6 GHz for 100\% nuclear polarization in GaAs
\cite{Paget:1977}. Changes in the average nuclear polarization
$\langle I \rangle / I$ within the QW can be measured directly as
changes in the precession frequency $\Delta \nu_L$.

\begin{figure}[b]\includegraphics{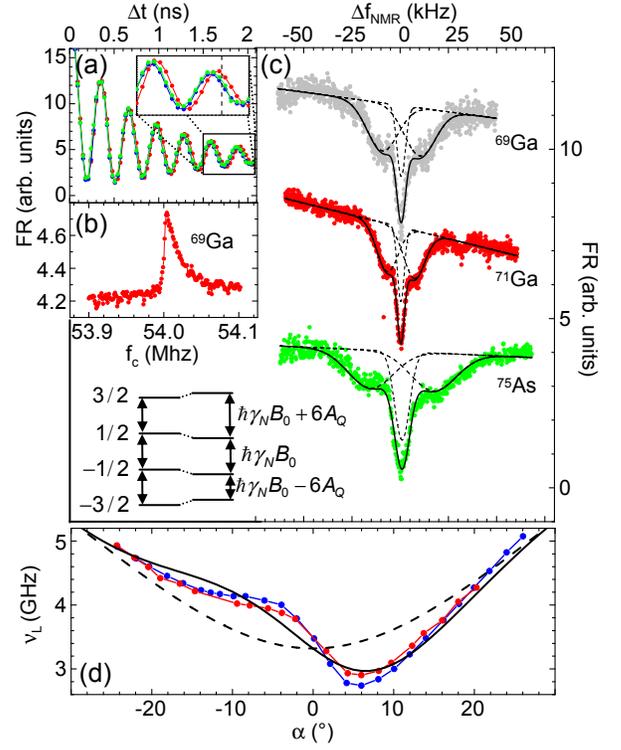}\caption{\label{fig2} 
    (Color) NMR detected by time-resolved FR. (a) FR data taken at
    $B_0 = 5.3019$ T with $B_1$ driven at 54.0000 MHz, 50.0400 MHz,
    and 53.9800 MHz for the red, blue, and green points respectively.
    Lines connect the data to guide the eye.  (b) FR shown as a
    function of $f_c$ for fixed $\Delta t = 1932$ ps as indicated by
    the dashed line in the inset to (a). (c) FR plotted as a function
    of detuning $\Delta f_{NMR}$ from the $^{69}$Ga resonance of
    52.9539 MHz at $B_0 = 5.2$ T, from the $^{71}$Ga resonance of
    67.2898 MHz at $B_0 = 5.2$ T, and from the $^{75}$As resonance of
    54.4992 MHz at $B_0 = 7.5$ T for the gray, red, and green points
    respectively.  Solid black lines are fits to the data while dashed
    lines show the three peaks included in those fits. A schematic
    diagram of the relevant level structure is included in the lower
    left. (d) $\nu_L$ shown as a function of $\alpha$ at $B_0 = 5.5$ T
    for no RF voltage applied to the coil and for -15 dBm applied at
    the $^{69}$Ga resonance at 56.0070 MHz for the blue and red points
    respectively. The solid black line is a fit to the angular
    dependence of $\nu_L$ in the presence of a non-zero $\langle I
    \rangle$.  The dashed black line shows the same dependence with
    $\langle I \rangle = 0$.  }\end{figure}

FR is plotted in Fig. \ref{fig2}a as a function of $\Delta t$ at $B_0
= 5.3019$ T with the coil driven continuously at a frequency $f_c$ set
to the $^{69}$Ga resonance at 50.0000 MHz and at two frequencies
slightly detuned from resonance. The inset clarifies the reduction of
$\nu_L$ for the resonant scan in which nuclear spin transitions
induced by $B_1$ decrease $\langle I \rangle$. Scans with an
off-resonant $f_c$, show the same $\langle I \rangle$ established by
DNP without any applied $B_1$. Fig. \ref{fig2}b shows FR data taken
under the same conditions as in Fig. \ref{fig2} while sweeping $f_c$
across the $^{69}$Ga resonance at a fixed $\Delta t = 1932$ ps. Here,
the resonant depolarization of $\langle I \rangle$ and the change in
$\nu_L$ appear as a peak in the FR signal. The asymmetry of the
resonance reflects the fast rate of the frequency sweep with respect
to the time required to polarize the nuclei $T_{DNP} \sim 90$ s. In
order to investigate the true form of the peak, $f_c$ is swept across
the full nuclear resonance in a time $T_{sweep} >> T_{DNP}$. This
condition is satisfied for the data shown in Fig. \ref{fig2}c where
the resonances due to the three isotopes present in the QW, $^{69}$Ga,
$^{71}$Ga, and $^{75}$As, appear at the expected frequencies. In
addition we observe satellite peaks for each resonance due to the
quadrupolar splitting $6 A_Q$. By fitting each resonance to a Gaussian
peak and two symmetric satellites, the splittings are measured to be
9.7 kHz, 7.0 kHz, and 16.3 kHz for the $^{69}$Ga, $^{71}$Ga, and
$^{75}$As isotopes.  These values are similar to previously reported
measurements and indicate the presence of a small amount of strain on
the crystal likely due the wax used in mounting the sample
\cite{Guerrier:1997}. The line-width (FWHM) of the main resonance is
2.6 kHz, 2.1 kHz, and 4.6 kHz for the $^{69}$Ga, $^{71}$Ga, and
$^{75}$As isotopes respectively. The line-widths of the satellite
peaks are broader at 9.4 kHz, 6.5 kHz, and 13.8 kHz probably because
of inhomogeneous strain in the sample.  As noted elsewhere
\cite{Guerrier:1997}, methods such as ours for accurately measuring
$A_Q$ are useful in the determination of built-in strain in
semiconductor heterostructures.

The dependence of $\nu_L$ on $\alpha$ is shown in Fig. \ref{fig2}d in
the case of no RF voltage applied to the transmission line and in the
case of -15 dBm applied at the $^{69}$Ga resonance $f_c = 5.0070$ MHz
for $B_0 = 5.5$ T. The solid black line is an angle dependence
calculated according to Salis et al. taking into account an
anisotropic $\hat{g}$ and a 9\% nuclear spin polarization.  The
calculation reproduces the qualitative features of the data and
confirms the dependence of DNP on $\alpha$. This analysis also leads
to the conclusion that the curve taken with the coil resonantly
depolarizing the $^{69}$Ga nuclei has a nuclear polarization of
$6-7$\%.  Since the natural abundance of $^{69}$Ga is $0.3$, we can
say that the RF coil is close to achieving full depolarization of the
resonant isotope within the QW.

A conventional ODNMR measurement was made in order to compare its
sensitivity to the FR-based scheme. In this case, a sample of bulk
semi-insulating (100) GaAs is used instead of a QW due to the lower
expected sensitivity of the photoluminescence (PL) technique. 5 mW of
circularly polarized light from a CW Ti:Sapphire laser tuned to 1.570
eV is focused to a 100 $\mu$m diameter spot on the sample surface.
For this experiment, the sample geometry is the same as shown in Fig.
\ref{fig1} with $\alpha = 20^{\circ}$.  Instead of collecting a
transmitted probe beam, here the polarization of the PL $\rho$ emitted
by the sample along the x-axis is measured using a 40 kHz
photo-elastic modulator, followed by a linear polarizer, and a
spectrometer with a photo-multiplier tube.  Emission from the
excitonic peak at 1.514 eV is collected at 5 K as a function of $B_0$.

\begin{figure}[b]\includegraphics{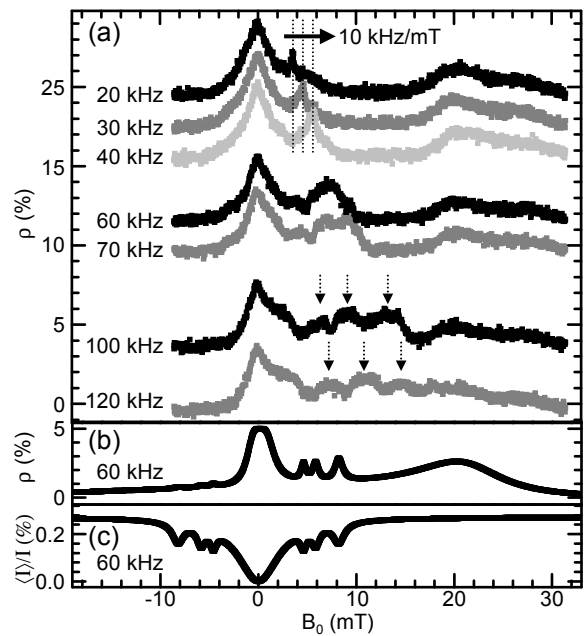}\caption{\label{fig3} 
    NMR detected by time-averaged PL polarization. (a) $\rho$ taken as
    $B_0$ is swept from -10 mT to 30 mT at 2 mT$/$min for different
    values of $f_c$. (b) A calculation of $\rho$ as a function of
    $B_0$ is shown along with the corresponding (c) dependence of
    $\langle I \rangle / I$ on $B_0$. The dip around $B_0 = 0$ T is
    due to the emergence of nuclear spin-spin coupling at low fields.
  }\end{figure}

Fig. \ref{fig3}a shows $\rho$ as a function of $B_0$ with $B_1$ driven
at $f_c$. The Hanle effect data shown here is typical of GaAs in the
presence of DNP \cite{OpticalOrientation:1984} and clearly illustrates
the resonant depolarization at the isotopic NMR frequencies. Electron
spin precession causes the time-averaged spin vector, and thus $\rho$,
to decrease in an increasing transverse magnetic field. By the same
reasoning, $\rho$ is sensitive to the effective transverse magnetic
field $B_n$ due to $\langle I \rangle$.  The broad peak seen around 20
mT in Fig.  \ref{fig3} is a result of $B_n$ directly opposing and
compensating $B_0$. The peaks shown to shift as a function of $f_c$
are due to a decrease in $B_n$ under the resonant depolarization of
$\langle I \rangle$. The 10 kHz$/$mT shift in the resonance for small
$f_c$ is close to the gyromagnetic ratio of the three relevant
isotopes. As $f_c$ increases, the splitting between resonances
increases until at $f_c = 100$ kHz and $f_c = 120$ kHz, the three
resonances are clearly distinguishable.

A calculation of the Hanle effect based on typical bulk GaAs
parameters and the three NMR resonances is shown in Fig. \ref{fig3}b.
There is good qualitative agreement between the model and the data
allowing us to estimate $\langle I \rangle / I \sim 0.25$\% as shown
in the dependence predicted by the model in Fig. \ref{fig3}c.  The
signal-to-noise in the data indicates that we are sensitive to changes
down to 0.05\%. Since the region from which we are collecting PL
contains $\sim 10^{16}$ nuclei, we estimate a sensitivity of $10^{12}$
nuclear spins for this ODNMR technique. In the FR measurement, which
was done in a QW, we probed many fewer nuclei, $\sim 10^{12}$. There
we could distinguish nuclear polarizations as small as $0.015$\%
corresponding to a sensitivity of $10^{8}$ nuclear spins.

In conclusion, ODNMR detected by time-resolved FR is an extremely
sensitive probe of nuclear polarization capable of resolving small
numbers of nuclear spins and distinguishing quadrupolar splittings in
the kHz range. It may find use in the determination of built-in strain
in GaAs heterostructures and provides an excellent way to perform
ODNMR measurements at high magnetic fields, impossible by conventional
techniques based on PL polarization.

\begin{acknowledgments}
  We thank Y. Ohno and H. Ohno for providing the samples, G. M.
  Steeves for his assistance in running the experiment, S. K.
  Buratto, and G. Salis for helpful discussions, and we acknowledge
  support from DARPA, ONR, and NSF.
\end{acknowledgments}


\begin{thebibliography}{24}
\bibitem{Lampel:1968} G. Lampel, Phys. Rev. Lett. \textbf{20}, 491
  (1968).
\bibitem{Barrett:1994} S. E. Barrett, R. Tycko, L. N. Pfeiffer, and K.
  W. West, Phys. Rev. Lett. \textbf{72}, 1368 (1994).
\bibitem{Ekimov:1972} A. I. Ekimov and V. I. Safarov, Zh. Eksp. Teor.
  Fiz. Pis'ma \textbf{15}, 257 (1972) [JETP Lett. \textbf{15}, 179
  (1972)].
\bibitem{Paget:1977} D. Paget, G. Lampel, B. Sapoval, and V. I.
  Safarov, Phys. Rev. B \textbf{15}, 5780 (1977).
\bibitem{Kalevich:1980} V. K. Kalevich, V. D. Kul'kov, and V. G.
  Fleisher, Sov. Phys. Solid State \textbf{22}, 703 (1980).
\bibitem{Kalevich:1981} V. K. Kalevich, V. D. Kul'kov, and V. G.
  Fleisher, Sov. Phys. Solid State \textbf{23}, 892 (1981).
\bibitem{Kalevich:1986} V. K. Kalevich, Sov. Phys. Solid State
  \textbf{28}, 1947 (1986).
\bibitem{Eickhoff:2002} M. Eickhoff, B. Lenzman, G. Flinn, and D.
  Suter, Phys. Rev. G \textbf{65}, 125301 (2002).
\bibitem{Paget:1982} D. Paget, Phys. Rev. B \textbf{25}, 4444 (1982).
  
\bibitem{Crooker:1997} S. A. Crooker, D. D. Awschalom, J. J. Baumberg,
  F. Flack, and N. Samarth, Phys. Rev. B \textbf{56}, 7574
  (1997).
\bibitem{Kikkawa:2000} J. M. Kikkawa and D. D. Awschalom, Science
  \textbf{287}, 473 (2000).
\bibitem{SalisPRL:2001} G. Salis, D. T. Fuchs, J. M. Kikkawa, D. D.
  Awschalom, Y. Ohno, and H. Ohno, Phys. Rev. Lett. \textbf{86}, 2677
  (2001).
\bibitem{SalisPRB:2001} G. Salis, D. D. Awschalom, Y. Ohno, and H.
  Ohno, Phys. Rev. B \textbf{64}, 195304 (2001).
\bibitem{Poggio:2003} M. Poggio, G. M. Steeves, R. C. Myers, Y. Kato,
  A. C. Gossard, and D. D. Awschalom, Phys. Rev. Lett. \textbf{91},
  207602 (2003).
\bibitem{Marie:2000} X. Marie, T. Amand, J. Barrau, P. Renucci, and P.
  Lejeune, Phys. Rev. B \textbf{61}, 11605 (2000).
\bibitem{Ohno:1999} Y. Ohno, R. Terauchi, T. Adachi, F. Matsukura, and
  H. Ohno, Phys. Rev. Lett. \textbf{83}, 4196 (1999).
\bibitem{Guerrier:1997} D. J. Guerrier and R. T. Harley, Appl. Phys.
  Lett. \textbf{70} 1739 (1997).
\bibitem{OpticalOrientation:1984} \textit{Optical Orientation}, edited
  by F. Meier and B. P. Zackharchenya (Elsevier, Amsterdam, 1984).
\end{thebibliography}
\end{document}